\documentclass[10pt]{article}

\usepackage{graphicx}
\usepackage{amsmath}
\usepackage{amssymb}
\usepackage{theorem}

\usepackage[T1]{fontenc}
\usepackage{times}

\theoremstyle{break}

\newtheorem{Def}{Definition}
\newtheorem{Theo}{Theorem}
\newtheorem{Prop}[Theo]{Proposition}

\newtheorem{Cor}{Corollary}[Theo]

\newenvironment{Proof}[1][Proof]{\paragraph{{#1}}}%
                {{\hfill\(\Box\)\\}}
                {{\hfill\(\Box\)\\}}

        {\paragraph{{#1}}\begin{list}{}{}\item
        }{\end{list}}

\newcommand{\cand}{\text{ and }}

\newcommand{\tuple}[1]{\ensuremath{\left\langle{#1}\right\rangle}}
\newcommand{\coll}[1]{\ensuremath{\left\{ {#1}\right\} }}
\newcommand{\pair}[2]{\ensuremath{\left\langle {#1}, {#2} \right\rangle}}

\newcommand{\paren}[1]{\ensuremath{\left( {#1} \right)}}

\newcommand{\set}[2]{\ensuremath{\left\{\left.#1\,\,\vphantom{#2}\right|\,#2\right\}}}
\newcommand{\fall}[1]{{\forall\,{#1},\ }}
\newcommand{\fexist}[1]{{\exists\,{#1}\,{:}\ }}

\newcommand{\mc}[1]{{\mathcal{#1}}}

\def\sas{\,\&\,}

\title{Representation Systems and Quantum Structures}
\author{Olivier Brunet \\ Leibniz - IMAG - University of Grenoble \\ 46, avenue Félix Viallet - 38031 GRENOBLE Cedex - France}
\begin{document}

\maketitle

\begin{abstract}
Two important classes of quantum structures, namely orthomodular posets and orthomodular lattices, can be characterized in a classical context, using notions like partial information and points of view. Using the formalism of representation systems, we show that these quantum structures can be obtained by expressing conditions on the existence of particular points of view, of particular ways to observe a system.
\end{abstract}

\section{Introduction}

The study of quantum structures such as orthomodular posets and orthomodular lattices constitutes an important part of the efforts to understand the relationship that exists between the quantum world and the classical, newtonian one. The traditional approach to this kind of study relies on the decomposition of an orthomodular structure into blocks, that is into maximal boolean subalgebras (one can refer to \cite{Ptak91Book}, \cite{Hughes89Book}, \cite{Svozil98Book}, or \cite{DallaChiara2001QuantumLogic}).


We present another approach, based on the decomposition of an orthomodular structure into complete boolean subalgebras. To this respect, we introduce representation systems which are an algebraic structure aimed at modelling partial knowledge about a system with an explicit notion of ``point of view'': if one considers a way to observe a system in a classical manner, it is natural to associate to this point of view a finite (or more generally, complete) boolean algebra whose elements correspond to partial knowledge about the state of the system, this partial knowledge following from information obtained from the considered point of view. In particular, the consideration of partial knowledge provides an intuitive justification for decomposing an orthomodular structure into complete boolean subalgebras.

We started this study in \cite{Brunet04IJTP} and in the present article, we show that under some conditions about the existence of particular points of view and about the way they relate to each other, the set of all partial descriptions of the system, regardless of their originating point of view, constitutes an orthomodular poset or an orthomodular lattice. This way, we provide a characterization of these quantum structures by means of purely classical notions such as that of point of view or of partial information.


In the next section, we introduce representation systems. Then, we focus on a restriction of these structures by demanding that each point of view is associated to a boolean algebra. In section 4, we show that our formalism can be used to define quantum structures by imposing conditions on the existence of adequate points of view. Finally, in section 5, we show that every orthomodular poset and orthomodular lattice can be obtained in this way.

\section{Representation Systems}

Representation systems \cite{Brunet02PhD,Brunet03JLC,Brunet04IJTP} are an algebraic structure whose purpose is to model partial knowledge about a system. They are based on two important related notions: points of view and partial information. A point of view corresponds to a way to ``observe'' the system (the verb {\em observe} is used here with its general meaning and not with its quantum acception) and gain information about its state. In particular, it might not be possible from a given point of view to totally describe the observed system. As a consequence, the information has to be considered in general as partial (i.e. is not sufficient to totally characterize the state of the system) in this context.

To each of these points of view, one can associate a poset whose elements represent partial descriptions of the state of the system, these partial descriptions corresponding to information obtainable from the considered point of view. This means in particular that a partial description associated to one point of view cannot in general be associated to another point of view. However, we assume that knowledge about the general structure of the system allows us to translate partial descriptions from one point of view to another, with the restriction that some information can be lost in the process. This assumption is formalized by what we call {\em transformation functions} in our formalism.

A detailled presentation of these structures can be found in \cite{Brunet03JLC} and in \cite{Brunet04IJTP}.

\begin{Def}[Representation System]
A {\em representation system} is a tuple $\mc S = \tuple{I, \coll{\pair{\mc P_i}{\leq_i}}_{i \in I}, \coll{f_{i|j}}_{i,j \in I}}$ where $I$ is a set of indices, where for every $i$ in $I$, $\pair{\mc P_i}{\leq_i}$ is a poset, and where the functions $\coll{f_{i|j} : \mc P_j \rightarrow \mc P_i}_{i,j \in I}$, called {\em transformation functions}, verify the following three properties:
\begin{align}
\forall {i \in I}, & \, f_{i|i} = {\rm id}_i & \text{Identity} \label{Eq:Identity} \\
\fall {i,j \in I} \forall {x,y \in \mc P_j}, & \, x \leq_j y \Rightarrow f_{i|j}(x) \leq_i f_{i|j}(y) & \text{Monotony} \label{Eq:Monotony} \\
\fall {i,j,k \in I} \forall {x \in \mc P_k}, & \, f_{i|k}(x) \leq_i f_{i|j} \circ f_{j|k}(x) & \text{Composition} \label{Eq:Composition}
\end{align}
\end{Def}

\subsubsection*{Example}

Consider an experiment with a firefly trapped in a box. This box is divided into 4 sectors (numbered from 1 to 4 in the figure below, on the left) and has an opaque division between sectors 2 and 4. At a given moment, two observers (X and Y) tell whether they see the light of the firefly, and in that case, in which half of the box they see it.

This situation can be modelled by means of a representation system with two points of view $X$ and $Y$, corresponding to the two observers. The corresponding posets are depicted below on the right. For instance, the elements of the $X$-poset are ``$\top$'' (an information-less description), ``Not seen'', ``Seen'', ``Left'' and ``Right'', depending whether the light of the firefly has been seen or not, and in the second case, in which half of the box it has been seen.

Moreover, in the same figure, the arrows depict the behaviour of the transformation functions. For instance, the arrow from ``Right'' to ``Down'' correspond to the equality $f_{Y|X}(\text{``Right''})=\text{``Down''}$ and means that if $X$ sees the light of the right half of the box, then the firefly is lit and is in sector 4, which corresponds, from $Y$'s point of view, to description ``Down''. One can note that we have only respresented the meaningful arrows, which are sufficient to entirely determine the transformation functions. Thus, one has $f_{Y|X}(\text{``Left''})=\text{``Seen''}$ and $f_{X|Y}(\text{``Up''})=\text{``}\top\text{''}$.

\bigskip

\noindent \begin{tabular}{cc}
\includegraphics[width=4cm]{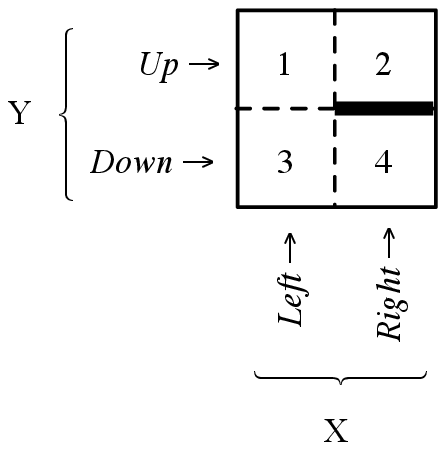} & \includegraphics[width=7cm]{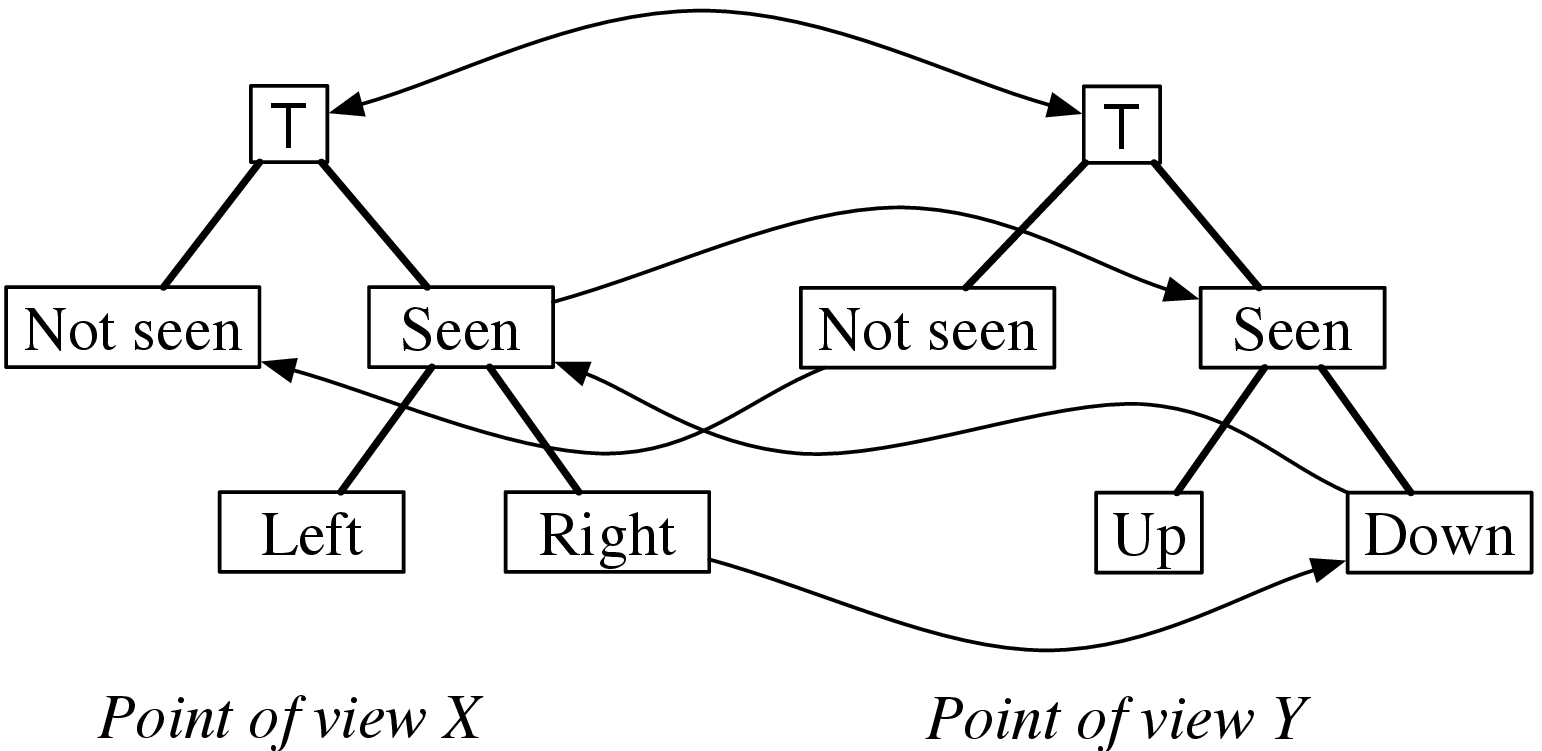} \\
\medskip Situation & Representation system
\end{tabular}

\subsubsection*{Sum of a Representation System}

Given a representation system, it is possible to merge the posets associated to the different points of view into a single poset. This way, one obtains a structure containing all the possible partial descriptions, regardless of the associated points of view.

\begin{Def}[Pre-Sum of a Representation System]
Let $\mc S = \tuple{I, \coll{\mc P_i}_{i \in I},\coll{f_{i|j}}_{i,j \in I}}$ be a representation system and define its pre-sum as the pair $\pair {\mc S_\star}{\leq_\star}$ where:
$$ \mc S_\star = \set {\pair i x}{i \in I \cand x \in \mc P_i} \quad \quad \quad \pair i x \leq_\star \pair j y \ \Leftrightarrow \,\ f_{j|i}(x) \leq_j y$$
\end{Def}

\begin{Prop}
The pre-sum $\pair{\mc S_\star}{\leq_\star}$ of a representation system $\mc S$ is a pre-ordered set, or equivalently $\leq_\star$ is a reflexive and transitive relation on $\mc S_\star$.
\end{Prop}

This pre-order induces an equivalence relation $\simeq_\star$ on $\mc S_\star$ by defining:
$$ \pair i x \simeq_\star \pair j y \ \Leftrightarrow \,\ \pair i x \leq_\star \pair j y \cand \pair j y \leq_\star \pair i x $$
Let $\pair i x_{\simeq_\star}$ denote the equivalence class of an element $\pair i x \in \mc S_\star$ with regards to $\simeq_\star$.
\begin{Def}[Sum of a Representation System]
Let $\mc S = \tuple{I, \coll{\mc P_i}_{i \in I},\coll{f_{i|j}}_{i,j \in I}}$ be a representation system and define its sum as the pair $\pair {\mc S_{\simeq_\star}}{\leq_{\simeq_\star}}$ where:
\begin{equation*}
\mc S_{\simeq_\star} = \set {\pair i x_{\simeq_\star}}{\pair i x \in \mc S_\star} \quad \quad \quad
\pair i x_{\simeq_\star} \leq_{\simeq_\star} \pair j y_{\simeq_\star} \ \Leftrightarrow \,\ \pair i x \leq_\star \pair j y
\end{equation*}
\end{Def}
\begin{Prop}
The sum $\mc S_{\simeq_\star}$ of a representation system $\mc S$ is a poset.
\end{Prop}
\begin{Prop}
For $i \in I$ and $x,y \in \mc P_i$, one has: $\pair i x_{\simeq_\star} \leq_{\simeq_\star} \pair i y_{\simeq_\star} \ \Leftrightarrow \,\ x \leq_i y$
\end{Prop}


It is possible to adapt the notion of point of view of a representation system to its sum by defining special closure operators on the sum. Let us first recall the definition of an upper closure operator.

\begin{Def}[Upper Closure Operator]
Given a poset $\pair {\mc P} \leq$, an {\em upper closure operator} on $\mc P$ is a monotonic function $ \rho : \mc P \rightarrow \mc P$ which verifies for all $x$:
\begin{align*}
\text{Idempotence: \quad} & \rho (\rho(x)) = \rho(x) & 
\text{Extension: \quad} & x \leq \rho(x)
\end{align*}
\end{Def}

For every $i \in I$, define a function $\rho_i : \mc S_{\simeq_\star} \rightarrow \mc S_{\simeq_\star}$ by: 
$$ \rho_i\paren{\pair j x_{\simeq_\star}} = \pair i {f_{i|j}(x)}_{\simeq_\star} $$

\begin{Prop}
Given a representation system $\mc S$, every $\rho_i$ is an upper closure operator on $\mc S_{\simeq_\star}$.
\end{Prop}
\begin{Proof}
{\em Extension} is shown as follows:
$$ \pair i x \leq_\star \pair j {f_{j|i}(x)} \Leftrightarrow f_{j|i}(x) \leq_j f_{j|i}(x) $$
{\em Idempotence} is a consequence of {\em Identity} (Eq. \ref{Eq:Identity}): $f_{i|i}(x)=x$ and {\em Monotony} of {\em Composition} (Eq. \ref{Eq:Composition}): \par \medskip
\hfill \hfill $ \pair i x \leq_\star \pair j y \Rightarrow f_{j|i}(x) \leq_j y \Rightarrow f_{k|i}(x) \leq_k f_{k|j}(y) $ \hfill
\end{Proof}

\noindent Intuitively, an element $a \in \mc S_{\simeq_\star}$ such that $a = \rho_i(a)$ corresponds to information that can be observed from point of view $i$. In general, one has $a \leq_{\simeq_\star} \rho_i(a)$, so that only a part of the information corresponding to $a$ can be observed from point of view $i$.

\section{Boolean Representation Systems}

A natural constraint which can be added to the formalism of representation systems is to assume that the poset associated to a given point of view forms a boolean algebra. This corresponds to the classical assumption of the ``newtonian'' world (as opposed to the quantum one) that knowledge behaves in the manner of classical logic.

By moreover adding conditions on transformation functions in order to take into account operations of boolean algebras (in particular, orthocomplementation and disjunction), we get the following definition:
\begin{Def}[Boolean Representation System]
A representation system $\mc S$ is {\em boolean} if and only if, using the usual notations:
\begin{enumerate}
\item Every poset $\mc P_i$ is a boolean algebra
\item The transformation functions verify:
\begin{align}
\fall {i,j \in \mc I} \forall {x,y \in \mc P_j}, & \, f_{i|j}(x \vee y) = f_{i|j}(x) \vee f_{i|j}(y) \label{Eq:SumJoin} \\
\fall {i,j \in \mc I} \fall {x \in \mc P_j} \forall {y \in \mc P_i}, & \, f_{i|j}(x) \leq_i y \Rightarrow f_{j|i}(y^\bot) \leq_j x^\bot \label{Eq:SumOrtho}
\end{align}
\end{enumerate}
\end{Def}

The following propositions illustrate some properties of the sum of a boolean representation system.

\begin{Prop} \label{Prop:SumNeg}
Given a boolean representation system $\mc S$, for every $\pair i x$ and $\pair j y$ in $\mc S_\star$, one has:
$$ {\pair i x} \leq_\star \pair j y \ \Rightarrow \,\ \pair j {y^\bot} \leq_\star \pair i {x^\bot} $$
\end{Prop}
\begin{Proof}
This is a direct consequence of equation \ref{Eq:SumOrtho}: \par \medskip
\hfill \hfill $\pair i x \leq_\star \pair j y \Rightarrow f_{j|i}(x) \leq_j y \Rightarrow f_{i|j}(y^\bot) \leq_i x^\bot \Rightarrow \pair j {y^\bot} \leq_\star \pair i {x^\bot}$ \hfill
\end{Proof}
\begin{Cor}
Given a boolean representation system $\mc S$, the operation ${\pair i x}_{\simeq_\star} \mapsto {\pair i {x^\bot}}_{\simeq_\star}$ is well-defined and constitues an orthocomplementation on $\mc S_{\simeq_\star}$.
\end{Cor}

\begin{Prop} \label{Prop:SumJoin}
Given a boolean representation system $\mc S$, for every $i \in \mc I$ and $x,y \in \mc P_i$, the join of ${\pair i x}_{\simeq_\star}$ and ${\pair i y}_{\simeq_\star}$ exists in $\mc S_{\simeq_\star}$ and is equal to ${\pair i {x \vee_i y}}_{\simeq_\star}$.
\end{Prop}
\begin{Proof}
First, one has ${\pair i x} \leq_\star {\pair i {x \vee_i y}}$ and a similar inequality for $\pair i y$. Conversely, suppose that one has ${\pair i x} \leq_\star {\pair j z}$ and ${\pair i y} \leq_\star {\pair j z}$. In that case, $f_{j|i}(x) \leq_j z$ and $f_{j|i}(y) \leq_j z$ so that as a consequence of equation \ref{Eq:SumJoin}, one has $f_{j|i}(x \vee y) \leq _i z$ which is equivalent to ${\pair i {x \vee y}} \leq_\star {\pair j z}$.
\end{Proof}

For the following proposition, let $\bot$ (resp. $\top$) denote the least (resp. greatest) element of a boolean algebra.

\begin{Prop}
Given a boolean representation system $\mc S$, its sum $\mc S_{\simeq_\star}$ is bounded and the least and greatest elements are given respectively by $ {\pair i \bot}_{\simeq_\star}$ and ${\pair i \top}_{\simeq_\star}$ for any $i \in \mc I$.
\end{Prop}

These results can be summarized in the following proposition:
\begin{Prop}
Given a {\em boolean representation system} $\mc S$, its sum $\mc S_{\simeq_\star}$ is a bounded orthoposet.
\end{Prop}

In terms of closure operators, if $\mc S$ is a boolean representation system then for $x, y \in \mc S_{\simeq_\star}$ and $i \in I$, it follows directly from propositions \ref{Prop:SumNeg} and \ref{Prop:SumJoin} that if $x = \rho_i(x)$, then $x^\bot = \rho_i(x^\bot)$ and if $x = \rho_i(x)$ and $y = \rho_i(y)$ then $x \vee y$ exists and verifies $x \vee y = \rho_i(x \vee y)$.

\section{Boolean Representation Systems and \\ Quantum Structures}

In the previous section, we have introduced boolean representation systems and shown that their sum is an orthoposet. We now study some conditions about the existence of appropriate points of view and characterize their sum.

\subsubsection*{Orthomodular Posets}

The first condition we introduce states that two elements $a$ and $b$ such that $a \leq b$ can be observed from a single point of view.

\begin{Prop} \label{Prop:CondOMP}
Let $\mc S$ be a boolean representation system such that:
\begin{equation}
\fall {a, b \in \mc S_{\simeq_\star}} \paren{\vphantom{a^b} a \leq_{\simeq_\star} b \Rightarrow \paren{\fexist {i \in I} a = \rho_i(a) \cand b = \rho_i(b)}} \label{Eq:OMPBRS}
\end{equation}
Then $\mc S_{\simeq_\star}$ is an orthomodular poset.
\end{Prop}
\begin{Proof}
This results from the fact that with the above condition, two elements verifying $a \leq_{\simeq_\star} b$ belong to a boolean subalgebra of $\mc S_{\simeq_\star}$. As a consequence, $a \vee b^\bot$ exists, and one has $ b = a \vee (b \wedge a^\bot) $.
\end{Proof}

\subsubsection*{Orthomodular Lattices}

The second condition we wish to study states that given two elements $a$ and $b$, there exists a  ``preferred'' point of view $i$ from which $a$ is observable and such that one can get as much information about $b$ as possible:
$$ \rho_i(a)=a \cand \fall j \paren{ \rho_j(a)=a \Rightarrow \rho_i(b) \leq \rho_j(b)} $$

But before this, we introduce a characterization of orthomodular lattices as orthomodular poset equipped with a particular binary operation \& (which can be shown to correspond to the Sasaki projection):
\begin{Prop} \label{Prop:SasakiOML}
Let $\mc P$ be an orthomodular poset equipped with a binary operation $\&$ which verifies:
\begin{align}
& \fall {x_1, x_2, y \in \mc P} x_1 \leq x_2 \Rightarrow x_1 \sas y \leq x_2 \sas y & \text{\&-Monotony} \label{Eq:LMon}\\
& \fall {x, y \in \mc P} x \sas y \leq y & \text{\&-Reduction} \label{Eq:RRed} \\
& \fall {x, y \in \mc P} x \leq y \Rightarrow x \sas y = x & \text{\&-Orthomodularity} \label{Eq:Ortho} \\
& \fall {x, y, z \in \mc P} x \sas y \leq z \Rightarrow z^\bot \sas y \leq x^\bot & \text{\&-Galois} \label{Eq:Galois}
\end{align}
Then $\mc P$ is an orthomodular lattice.
\end{Prop}
\begin{Proof}
Since $\mc P$ is an orthomodular poset, one only needs to show that it is also a lattice. For this, define a binary operation $\barwedge$ as $x \barwedge y = \paren{x^\bot \sas y}^\bot \sas y$ and let us show that $x \barwedge y$ is the meet of $x$ and $y$. First, it is clear from \&-Reduction that $x \barwedge y \leq y$. Moreover, one has $x \barwedge y \leq x$ since $x^\bot \sas y \leq x^\bot \sas y$ implies $\paren{x^\bot \sas y}^\bot \sas y \leq x$ using \&-Galois.

Finally, let $z$ be in $\mc P$ such that $z \leq x$ and $z \leq y$. One has:
\begin{align*}
& z \leq x & \text{Hypothesis} \\
\Rightarrow \,\ & z \sas y \leq x & \text{Hypothesis and \&-Orthomodularity} \\
\Rightarrow \,\ & x^\bot \sas y \leq z^\bot & \text{\&-Galois} \\
\Rightarrow \,\ & z \leq \paren{x^\bot \sas y}^\bot \\
\Rightarrow \,\ & z \sas y \leq \paren{x^\bot \sas y}^\bot \sas y & \text{\&-Monotony} \\
\Rightarrow \,\ & z \leq \paren{x^\bot \sas y}^\bot \sas y & \text{Hypothesis and \&-Orthomodularity}
\end{align*}
This means that is $z \leq x$ and $z \leq y$, then $z \leq x \barwedge y$, and finishes the proof that $x \barwedge y$ is the meet of $x$ and $y$.
\end{Proof}

\begin{Prop} \label{Prop:CondOML}
Let $\mc S$ be a boolean representation system such that equation \ref{Eq:OMPBRS} holds and that one has:
\begin{equation}
\fall {a,b \in \mc S_{\simeq_\star}} \fexist  {i \in I} \left\{\begin{array}{l} a = \rho_i(a) \cand \\ \fall j \paren{a = \rho_j(a) \Rightarrow \rho_i(b) \leq_{\simeq_\star} \rho_j(b)} \end{array}\right. \label{Eq:OMLBRS}
\end{equation}
Then $\mc S_{\simeq_\star}$ is an orthomodular lattice.
\end{Prop}

This condition corresponds to the fact that given two elements $a$ and $b$, there exists a least element $c$ compatible with $a$ such that $b \leq c$.
\begin{Proof}
Since equation \ref{Eq:OMPBRS} holds, $\mc S_{\simeq_\star}$ is an orthomodular poset, as it follows from proposition \ref{Prop:CondOMP}. As a consequence, it suffices to exhibit a binary operation as $\&$ in proposition \ref{Prop:SasakiOML}. For this, given $a$ and $b$ in $\mc S_{\simeq_\star}$, we define $a \sas b$ as $\rho_i(a) \wedge b$ with $i$ such that:
$$ \rho_i(b)=b \quad \cand \quad \fall {j \in I} \paren{\rho_j(b)=b \Rightarrow \rho_i(a) \leq_{\simeq_\star} \rho_j(a)} $$
We show that this operator verifies the properties given in equations \ref{Eq:LMon}-\ref{Eq:Galois}.
\begin{itemize}
\item For \&-Monotony (Eq. \ref{Eq:LMon}), let $a_1, a_2$ be in $\mc S_{\simeq_\star}$ such that $a_1 \leq_{\simeq_\star} a_2$, and let $i$ and $j$ be in $I$ such that $a_1 \sas b = \rho_i(a_1) \wedge b$ and $a_2 \sas b = \rho_j(a_2) \wedge b$. From the choice of $i$ and the monotony of $\rho_j$, one has:
$ \rho_i(a_1) \leq_{\simeq_\star} \rho_j(a_1) \leq_{\simeq_\star} \rho_j(a_2) $
so that $a_1 \sas b \leq a_2 \sas b$.
\item For \&-Reduction (Eq. \ref{Eq:RRed}), it is obvious that $a \sas b \leq b$ from its definition.
\item For \&-Orthomodularity (Eq. \ref{Eq:Ortho}), if $a \leq_{\simeq_\star} b$, then there exists an index $i \in I$ such that $a = \rho_i(a)$ and $b = \rho_i(b)$. As a consequence, one has $a \sas b = a \wedge b = a$.
\item Finally, for \&-Galois (Eq. \ref{Eq:Galois}), suppose that $a \sas b \leq_{\simeq_\star} c$. This means, with $a \sas b = \rho_i(a) \wedge b$, that $\rho_i(a) \wedge b \leq_{\simeq_\star} c$. As a consequence, $c^\bot \leq_{\simeq_\star} (\rho_i(a))^\bot \vee b^\bot$.

Now, following equation \ref{Eq:OMLBRS}, introduce $j \in I$ such that $c^\bot \sas b = \rho_j(c^\bot) \wedge b$, $\rho_j(c^\bot) \leq \rho_i(c^\bot)$ and $\rho_j(b)=b$. Since $\rho_i(b)=b$ and $\rho_i\paren{(\rho_i(a))^\bot \vee b^\bot}=(\rho_i(a))^\bot \vee b^\bot$, it follows that $\rho_j(c^\bot) \leq_{\simeq_\star} (\rho_i(a))^\bot \vee b^\bot$. Thus, one can write:
$$ \rho_j(c^\bot) \wedge b \leq \paren{(\rho_i(a))^\bot \vee b^\bot} \wedge b \leq (\rho_i(a))^\bot \wedge b \leq (\rho_i(a))^\bot \leq a^\bot $$
\end{itemize}
Thus, we have shown that $\mc S_{\simeq_\star}$ is both an orthomodular poset and a lattice.
\end{Proof}

\section{Representation of Quantum Structures}

The results in the previous section show that orthomodular posets and orthomodular lattices arise naturally in a context of partial representation of knowledge, where there exists a ``rich'' enough collection of points of view.
We now present the converse result, which states that these structures can always be obtained as the sum of boolean representation systems.

\ 

Let $\mc P$ be a bounded orthoposet, and define $I_{\mc P}$ as the set of complete boolean subalgebras of $\mc P$. Moreover, for all $\mc B \in I_{\mc P}$, define $\rho_{\mc B} : \mc P \rightarrow \mc B$ as:
$$ \rho_{\mc B}(x) = \bigwedge \set{y \in \mc B}{x \leq y} $$
Finally, for $\mc B,\mc B' \in I_{\mc P}$, let $f_{\mc B|\mc B'}$ denote the restriction of $\rho_{\mc B}$ to $\mc B'$.
\begin{Prop}
The tuple $\mc S_{\mc P} = \tuple{I_{\mc P}, \coll{\mc B}_{\mc B \in I_{\mc P}}, \coll{f_{\mc B|\mc B'}}_{\mc B, \mc B' \in I_{\mc P}}}$ is a boolean representation system.
\end{Prop}
\begin{Proof}
One just needs to prove that the transformation functions $\coll{f_{\mc B|\mc B'}}$ actually verify {\em Monotony}, {\em Idempotence} and {\em Composition}. {\em Monotony} and {\em Idempotence} directly follow from their definition. Concerning {\em Composition}, one has:
\begin{align*}
\set{y \in \mc B}{\rho_{\mc B'|\mc B''}(x) \leq y}
& = \set{y \in \mc B}{\bigwedge \set{z \in \mc B'}{x \leq z} \leq y} \\
& \subseteq \set{y \in \mc B}{x \leq y} \\
\smallskip \\
\text{so that \quad} \bigwedge \set{y \in \mc B}{x \leq y} & \leq \bigwedge \set{y \in \mc B}{\rho_{\mc B'|\mc B''}(x) \leq y}
\end{align*}
The last inequality is equivalent to $\rho_{\mc B|\mc B''}(x) \leq \rho_{\mc B|\mc B'} \circ \rho_{\mc B'|\mc B''}(x)$.
\end{Proof}

\begin{Prop} \label{Prop:SumOrtho}
The sum $\paren{\mc S_{\mc P}}_{\simeq_\star}$ is isomorphic to $\mc P$.
\end{Prop}
\begin{Proof}
Let $\pair {\mc B_x} x$ and $\pair {\mc B_y} y$ be two elements of $\paren{\mc S_{\mc P}}_\star$. If $\pair {\mc B_x} x \leq_\star \pair {\mc B_y} y$, then $\rho_{\mc B_y}(x) \leq y$ which implies that $x \leq y$. Thus, the elements of $\paren{\mc S_{\mc P}}_{\simeq_\star}$ are of the form $\set{\pair{\mc B}x}{\mc B \in I_{\mc P}, x \in \mc B}$ and can be put in a one-to-one correspondence with $x$.

It is easy to verify that this bijection preserves both the partial order relation and the orthocomplementation.
\end{Proof}
\begin{Prop}
All bounded orthoposets are isomorphic to the sum of a boolean representation system.
\end{Prop}

The notion of compatibility in the field of orthomodular structures can be easily expressed in our approach: two elements $a,b$ in $\mc P$ are compatible if and only if $\fexist{\mc B \in I_{\mc P}} \coll{a,b} \subseteq \mc B$.
\begin{Prop}
Every orthomodular poset is isomorphic to the sum of a boolean representation system which verifies:
\begin{equation*}
\fall {a, b \in \mc S_{\simeq_\star}} a \leq_{\simeq_\star} b \Rightarrow \fexist {i \in I} a = \rho_i(a) \cand b = \rho_i(b) \end{equation*}
\end{Prop}
\begin{Proof}
Two comparable elements $a \leq b$ of an orthomodular poset are compatible.
\end{Proof}

It should be remarked that the condition in this proposition is exactly equation \ref{Eq:OMPBRS} used in proposition \ref{Prop:CondOMP}.

\begin{Prop}
Every orthomodular lattice is isomorphic to the sum of a boolean representation system which verifies:
\begin{gather*}
\fall {a, b \in \mc S_{\simeq_\star}} a \leq_{\simeq_\star} b \Rightarrow \fexist {i \in I} a = \rho_i(a) \cand b = \rho_i(b) \\
\fall {a,b \in \mc S_{\simeq_\star}} \fexist  {i \in I} \left\{\begin{array}{l} a = \rho_i(a) \cand \\ \fall j \paren{a = \rho_j(a) \Rightarrow \rho_i(b) \leq_{\simeq_\star} \rho_j(b)} \end{array}\right.
\end{gather*}
\end{Prop}
\begin{Proof}
The second condition comes from the fact that $a$ and $(a \vee b) \wedge (a \vee b^\bot)$ are compatible, and that any element $c$ compatible with $a$ and such that $b \leq c$ verifies $(a \vee b) \wedge (a \vee b^\bot) \leq c$.
\end{Proof}

The second condition here is exactly equation \ref{Eq:OMLBRS} used in proposition \ref{Prop:CondOML}.

\section{Conclusion}

In this article, we have presented the notion of representation system and of boolean representation system which is designed to model partial knowledge about a system using several points of view, in such a way that each point of view corresponds to a classical observation of the system. By expressing conditions about the existence of particular points of view, we have shown that it is possible to characterize and represent quantum structures such as orthomodular posets and orthomodular lattices using these structures which are based on classical notions only.

\bibliographystyle{apalike}

\end{document}